\pdfoutput=1
\documentclass[aps,prd,twocolumn,superscriptaddress,tightenlines,nofootinbib]{revtex4}
\usepackage{graphicx}
\usepackage{epsfig}
\usepackage{bm}
\usepackage{latexsym,amssymb,amsmath,amsfonts,amssymb,txfonts,pxfonts,wasysym,float}
\usepackage{color}

\newcommand{\beq}[1]{\begin{equation}\label{#1}}
\newcommand{\eeq}{\end{equation}}
\newcommand{\bea}[1]{\begin{eqnarray} \label{#1}}
\newcommand{\eea}{\end{eqnarray}}
\newcommand{\ba}{\begin{array}}
\newcommand{\ea}{\end{array}}

\def\be{\begin{equation}}
\def\ee{\end{equation}}
\def\gs{\mathrel{
   \rlap{\raise 0.511ex \hbox{$>$}}{\lower 0.511ex \hbox{$\sim$}}}}
\def\ls{\mathrel{
   \rlap{\raise 0.511ex \hbox{$<$}}{\lower 0.511ex \hbox{$\sim$}}}}

\def\ie{i.e.\ }

\newcommand{\postscript}[2]{\setlength{\epsfxsize}{#2\hsize}
   \centerline{\epsfbox{#1}}}

\newcommand{\comment}[1]{}

\usepackage[usenames,dvipsnames]{xcolor}
\definecolor{orange}{cmyk}{0,0.5,1,0}
\definecolor{rossoCP3}{cmyk}{0,.88,.77,.40}
\definecolor{graa}{rgb}{0.8,0.8,0.8}
\definecolor{blaa}{rgb}{0.2,0.2,0.6}

\begin{document}

\title{\color{rossoCP3}{Cosmic Mass Spectrometer
}}

\author{Luis A. Anchordoqui}
\affiliation{Department of Physics \& Astronomy,  Lehman College, City University of
  New York, NY 10468, USA}
\affiliation{Department of Physics,
 Graduate Center, City University
  of New York,  NY 10016, USA}
\affiliation{Department of Astrophysics,
 American Museum of Natural History, NY
 10024, USA}

\author{Vernon Barger}
\affiliation{Department of Physics, University of Wisconsin, Madison,
WI 53706, USA}

\author{Thomas J. Weiler}
\affiliation{Department of Physics \& Astronomy, Vanderbilt University, Nashville, TN 37235, USA}

\begin{abstract}
  \noindent We argue that if ultrahigh-energy ($E \agt 10^{10}~{\rm
    GeV}$) cosmic rays are heavy nuclei (as indicated by existing
  data), then the pointing of cosmic rays to their nearest
  extragalactic sources is expected for $10^{10.6} \alt E/{\rm GeV}
  \alt 10^{11}$.  This is because for a nucleus of charge $Ze$ and
  baryon number $A$, the bending of the cosmic ray decreases as $Z/E$
  with rising energy, so that pointing to nearby sources becomes
  possible in this particular energy range.  In addition, the maximum energy of acceleration
  capability of the sources grows linearly in $Z$, while the energy
  loss per distance traveled decreases with increasing $A$.  Each of
  these two points tend to favor heavy nuclei at the highest energies.
  The traditional bi-dimensional analyses, which simultaneously
  reproduce Auger data on the spectrum and nuclear composition, may
  not be capable of incorporating the relative importance of all these
  phenomena.  In this paper we propose a multi-dimensional
  reconstruction of the individual emission spectra (in $E$,
  direction, and cross-correlation with nearby putative sources) to
  study the hypothesis that primaries are heavy nuclei subject to GZK
  photo-disintegration, and to determine the nature of the
  extragalactic sources.  More specifically, we propose to combine
  information on nuclear composition and arrival direction to
  associate a potential clustering of events with a 3-dimensional
  position in the sky. Actually, both the source distance and maximum
  emission energy can be obtained through a multi-parameter likelihood
  analysis to accommodate the observed nuclear composition of each
  individual event in the cluster.  We show that one can track the
  level of GZK interactions on an statistical basis by comparing the
  maximum energy at the source of each cluster. We also show that
  nucleus-emitting-sources exhibit a {\it cepa stratis} structure on Earth
  which could be pealed off by future space-missions, such as POEMMA.
  Finally, we demonstrate that metal-rich starburst galaxies are
  highly-plausible candidate sources, and we use them as an explicit
  example of our proposed multi-dimensional analysis.
\end{abstract}

\maketitle

\section{Introduction}
\label{intro}

The most important result so far from the present generation of
ultrahigh-energy cosmic ray (UHECR) observatories is the conclusive
evidence that the cosmic ray (CR) flux drops precipitously for
energies beyond $E \approx 10^{10.6}~{\rm GeV}$.  The discovery of
this suppression was first reported by the HiRes~\cite{Abbasi:2007sv}
and Auger collaborations~\cite{Abraham:2008ru}, and later confirmed by
the Telescope Array (TA)~\cite{AbuZayyad:2012ru}.\footnote{Evidence
  for a suppression in the spectrum, from data of first generation
  UHECR experiments, was pointed out in~\cite{Bahcall:2002wi}.} By now
(in Auger data) the suppression has reached a statistical significance
of more than $20\sigma$~\cite{Abraham:2010mj}.

There are two competing classes of models to explain the observed
suppression.  The competing models are the Greisen, Zatsepin, and
Kuz'min (GZK) effect due to the CR interaction with the cosmic
microwave background (CMB)~\cite{Greisen:1966jv,Zatsepin:1966jv}, and
the {\sl disappointing} model~\cite{Aloisio:2009sj} wherein it is
postulated that the ``end-of steam'' for cosmic accelerators is
coincidentally near the putative GZK cutoff, with the exact energy
cutoff determined by data.  Since both models accommodate the same
rate in the mean, it is a challenge to discriminate between them.  

In this paper we re-examine the GZK interactions in the nearby
universe (where the models differ most) and discuss a method that can
be used to discriminate between the two classes of models.  Indeed,
model discrimination becomes feasible by analyzing UHECR events beyond
the onset of the suppression. Throughout we will refer to these events
as trans-GZK events. We show that when information on nuclear
composition is combined with the distribution of arrival directions it
is possible to elaborate a concrete mapping of clustered trans-GZK
events, which can isolate the source location in 3-dimensions. To
positively associate a potential cluster with a 3-dimensional position
in the sky, we need information on the nuclear composition of each
event in the cluster. This is because each nuclear species exhibits
different propagation characteristics, which arrange a natural {\it
  mass spectrometer} in the local (distance $\alt 50~{\rm Mpc}$)
universe. At the same time, we can determine the maximum CR energy of
the source producing the cluster, which together with the propagation
distance controls the level of GZK interactions. As an illustrative
example, we invoke starburst galaxies as the sources of UHECRs.

GZK noted that the extragalactic UHECR flux could be dominated either
by protons or  nuclei, with the GZK effect driven by photo-pion
production and photo-disintegration,
respectively~\cite{Greisen:1966jv,Zatsepin:1966jv}.  For heavy nuclei,
photo-disintegration on the cosmic infrared (IR) background is also
relevant in modeling the high-energy tail of the
spectrum~\cite{Puget:1976nz}.  If UHECRs are protons then there is
only one {\it visible} effect of the GZK interactions, which is the
suppression of the spectrum around $10^{10.6}~{\rm GeV}$. However, as
we advanced above and demonstrate below, if UHECR are  nuclei
there is more to observe than just the suppression of the spectrum.

The atmospheric depth at which the longitudinal development of a CR
shower has its maximum, $X_{\rm max}$, and the number of muons $N_\mu$
reaching ground level are the most powerful observables to determine
the UHECR nuclear composition~\cite{Anchordoqui:2004xb}. The Auger
high-quality, high-statistics data, when interpreted with existing
hadronic event generators, exhibit a strong likelihood for a
composition that becomes gradually heavier with increasing energy,
beginning around $10^{9.7}~{\rm
  GeV}$~\cite{Abraham:2010yv,Aab:2014dua,Aab:2014kda,Aab:2014aea,Aab:2016htd}.
Within uncertainties, the data from TA are consistent with these
findings~\cite{Abbasi:2014sfa, Abbasi:2015xga}.  

In addition, TA has
observed a statistically significant excess in cosmic rays with
energies above $57~{\rm EeV}$ in a region of the sky spanning about
$20^\circ$, centered on equatorial coordinates R.A. = $146.7^\circ$,
Dec. = $43.2^\circ$~\cite{Abbasi:2014lda}. This is colloquially
referred to as the ``TA hot spot.'' The chance probability of this hot
spot in an isotropic CR sky was calculated to be $p_{\rm TA} = 3.7
\times 10^{-4}$ ($3.4\sigma$)~\cite{Kawata:2015whq}.  The absence of a
concentration of nearby sources in this region of the sky corroborates
other experimental evidence for UHECR nuclei, in that a few local
sources within the GZK sphere can produce the hot spot through
significant deflection and translation (proportional to the nucleus
charge $Ze$) in the
extragalactic and Galactic magnetic fields.  More recently, this
picture has been further supported by Auger data, which revealed an
intermediate-scale anisotropy, with statistical significance of
$4\sigma$~\cite{Aab:2017njo,ObservatoryMichaelUngerforthePierreAuger:2017fhr}.\footnote{Throughout ``intermediate'' denotes angular
  scales larger than the angular resolution of the Auger array, which
  is about $1^\circ$, and
  smaller than large-scale patterns, i.e., below about  $45^\circ$.}

If the highest-energy primary CRs are dominated by heavy nuclei, there
are important implications for the astrophysics of the sources.  For
example, a trend toward heavier composition could reflect the endpoint
of cosmic acceleration.  The acceleration of primaries is proportional
to $Z$, so heavy nuclei can be expected to dominate the composition
near the end of the spectrum (with $E_{\max}$ coincidentally falling
off near the expected GZK cutoff region~\cite{Aloisio:2009sj}).  In
such a model, the suppression would constitute an imprint of the
accelerator characteristics rather than energy loss in transit.  Such
a model has been termed the {\it disappointing}
model~\cite{Aloisio:2009sj}, in that no physics beyond acceleration in
sources is invoked. Suppressions due to source end points and due to
GZK losses are simultaneously possible.

The general idea behind our method of discrimination is summarized in the
following axioms (the first two arising from the GZK effect):
\begin{itemize}
\item The higher the energy the heavier the nuclear species. 
\item The higher the energy the smaller the number of apparent sources, because 
nuclei lose energy roughly proportional to their distance of travel, thus favoring the few nearby sources.
\item The higher the energy of a cosmic ray proton the smaller the
  bending on the magnetic field and therefore the smaller the angle
  between the incident CR direction at Earth and the line-of-sight to
  the true source. 
\item The higher the energy of a cosmic ray nuclei the bending decreases as $Z/E$.
\end{itemize}

The collection of spectral and anisotropic features and nuclear composition
observables undoubtedly reflect physically interesting phenomena,
including source distribution(s), emission properties, and propagation effects. 

A common approach to interpreting spectral features and nuclear
composition is to develop some hypothesis about source properties and,
using either analytic or Monte Carlo methods, to predict the mean
spectrum and the average nuclear composition expected at Earth.  As
our knowledge of source distributions and properties is limited, it is
common practice to assume spatially homogeneous and isotropic UHECR
emissions.  A mean spectrum and average nuclear composition are then
computed, based on this assumption.  If, following Auger and TA data, the
composition is taken to be mixed, then the simultaneous fit to the
spectrum and composition (e.g., $X_{\rm max}$ distribution) imposes
severe constraints on model parameters: {\it (i)} hard source spectra
and {\it (ii)}~``low'' energy cutoff, of order $10^{9.7} \, Z~{\rm
  GeV}$~\cite{Aloisio:2013hya,Unger:2015laa,Aab:2016zth}. We note in
passing that the constraint on the spectral index can be relaxed by
considering a negative source evolution with
redshift~\cite{Taylor:2015rla}, but the assumption of softer source
spectra leaves the energy cutoff unaltered (see e.g. Fig.~11
in~\cite{Unger:2015laa}). Now, a robust argument can be advanced: for the source
parameters given above,
it is extremely challenging to distinguish between the two classes of
models. The rationale will be given by example. At first sight it will
appear that if a substantial fraction of the trans-GZK events are
oxygen or lighter nuclei, then GZK interactions must be at play, as
the source maximum energy for oxygen would be $10^{10.6}~{\rm 
  GeV}$. However, we note that the earthly maximum energy for an
oxygen nucleus which is produced as a surviving fragment of a heavier
nucleus during propagation is roughly the same; say, for an iron
nucleus the source maximum energy would be $E_{\rm max} \sim
10^{11.2}~{\rm GeV}$, and so after losing 40 nucleons the energy of
the oxygen would
also be close to $10^{10.6}~{\rm GeV}$. This would make the two
classes of models almost indistinguishable.

The line of argumentation given above is of course limited to the
assumption of an homogeneous source distribution.  In reality this
assumption cannot be correct, especially at the highest energies where
the GZK effect severely limits the number of sources visible to us. We
can quantify the possible deviation from the mean prediction based on
the knowledge we do have on the source density and the pointing
distances to the closest source populations. This fluctuation about
the mean has been referred to as the ensemble
fluctuation~\cite{Ahlers:2012az}.  The local fluctuations will present
their strongest effect for the most energetic cosmic rays, due to the
limited propagation distance in the cosmic radiation background (see
e.g. Fig.~2 in~\cite{Ahlers:2013zxa}).  In other words, the strongest
effect is a local phenomenon, unique to our local Galactic geometry,
and a manifestation of the {\it cosmic variance}. 

As shown in~\cite{Taylor:2011ta}, the UHECR flux of sources beyond
100~Mpc can be well approximated by a homogeneous distribution of
sources. However, the finite (non-zero) distance to the nearest source
of UHECR nuclei leads to a breakdown of the homogeneous source
distribution spectrum result at the highest energies.  The combined
fit to the spectrum and $X_{\rm max}$ distribution considering a
non-zero distance to the first source leads to a maximum injection
energy in the range $10^{11.5} < E_{\rm max}/{\rm GeV} < 10^{12}$,
depending on the nuclear species. We
arrive then at the fifth axiom of our method of discrimination:
\begin{itemize}
\item The discreteness of nearby UHECR emitters must be
considered. 
\end{itemize}

Putting all this together we come up with a new analysis technique to
discriminate between the two classes of models. Namely, we propose to
combine information on nuclear composition and arrival direction to
associate a potential clustering of trans-GZK events with a
3-dimensional position in the sky. For a given cluster, the distance
to the source is determined through a multi-parameter fit to the
observed nuclear composition of each individual event, in conjunction
with possible GZK energy losses. Model discrimination can be done on
an statistical basis by comparing the best fit parameters of each
individual cluster.\footnote{It is noteworthy that the actual
  observation of the GZK effect would provide strong constraints on
  Lorentz invariant breaking effects. This is because if Lorentz
  invariance is broken in the form of non-standard dispersion
  relations, then absorption and energy loss
  processes for UHECR interactions would be modified. See e.g.~\cite{Coleman:1998ti}  for
  interactions of UHECR protons with the CMB.}

The rest of the paper is organized as follows. In Sec.~\ref{losses} we
re-examine the interactions of UHECR nuclei on the pervasive sea of
microwave and infrared radiation filling the universe.  We show that,
for $10^{10.5} \alt E/{\rm GeV} \alt􏰌 10^{11.5}$ and propagation
distances $\alt􏰌 50~{\rm Mpc}$, a fully analytic treatment of the
energy losses that UHECR nuclei suffer {\it en route} to Earth becomes
feasible. This is because the effects of pair and photo-meson
production can be safely neglected and we need to only consider a
single energy loss mechanism: photo-disintegration. Most importantly,
we point out that we can use information on the nuclear composition to
pin down the origin of any potential cluster of trans-GZK events
appearing in the distribution of arrival directions. We also comment
on the experimental sensitivity of such analysis for existing and
future UHECR observatories. In Sec.~\ref{cepastratis} we confront the
information on magnetic field deflections contained in our third and
fourth axioms to develop a new method to discriminate between
accelerators of UHECR protons and sources of UHECR nuclei. In
Sec.~\ref{starburst} we particularize our discussion to starburst
galaxies. We first revise a two-step acceleration model presented
elsewhere~\cite{Anchordoqui:1999cu} to accommodate the new needs of
source spectra that can reproduce Auger data.  After that we derive
accurate predictions for the earthly nuclear composition that can be
confronted with experimental data on an event-by-event basis. We show
that existing data are consistent with the characteristics of
nucleus-emitting-sources delineated in Sec.~\ref{cepastratis}. The
paper wraps up with some conclusions presented in Sec.~\ref{summary}.

\section{Energy Loss as a Trace of Source Distance}
\label{losses}

The relevant mechanisms for the GZK energy loss that extremely
high-energy nuclei are expected to suffer on the way to Earth are:
pair production in the field of the nucleus, photo-disintegration, and
meson photo-production.  In the nucleus rest-frame, pair production
has a threshold at $\sim 1~{\rm MeV}$, photo-disintegration is
particularly important at the peak of the giant dipole resonance (GDR)
that corresponds to photon energies of 15 to 25~MeV, and photo-meson
production has a threshold energy of $\sim 145~{\rm
  MeV}$.\footnote{For photo-disintegration, the averaged fractional
  energy loss equals the fractional loss in baryon number of the
  nucleus. During the photo-disintegration process the Lorentz factor
  of the nucleus is conserved, unlike the cases of pair production and
  photo-meson production processes, which involve the creation of new
  particles that carry off energy.}  For $10^{10.5} \alt E/{\rm GeV}
\alt 10^{11.5}$ and propagation distances $\alt 50~{\rm Mpc}$, the
effect of pair and photo-meson production can be safely
neglected~\cite{Blumenthal:1970nn}, \ie at the high-energy end of the
heavy-nuclei spectrum, the photo-disintegration process dominates the
energy losses.\footnote{The inelasticity of $(e^+e^-$) pair production
  is very low ($\approx m_e/m_p$, for $Z=1$) so that the energy loss
  of UHECRs is gradual. The characteristic lifetime for energy loss
  for this process at energies $\agt 10^{10}~{\rm GeV}$ is $\tau = E/
  (dE/dt) \approx 5 \times 10^{9}~{\rm yr}$~\cite{Aharonian:1994nn}.
  For a nucleus, the energy loss rate is $Z^2/A$ times higher than for
  a proton of the same Lorentz factor~\cite{Chodorowski}.} With this
dominance, we now exploit a complete analytic treatment of the GZK
effect.

The interaction time $\tau^{\rm int}_A$ for a highly
relativistic nucleus with energy $E=\gamma A m_p$ (where $\gamma$ is
the Lorentz factor) propagating through an isotropic photon background
with energy $\varepsilon$ and spectrum $dn(\varepsilon)/d\varepsilon$,
is~\cite{Stecker:1969fw}
\begin{equation}
\frac{1}{\tau_{\rm int}} =
\frac{c}{2} \, \int_0^{\infty} \frac{1}{\gamma^2 \varepsilon^2} \
\frac{d n(\varepsilon)}{d \varepsilon}
\, d\varepsilon \, \int_0^{2\gamma \varepsilon} \varepsilon' \,
\sigma_A(\varepsilon') \, d\varepsilon' \, ,
\label{rate}
\end{equation}
where $\sigma_A (\varepsilon')$ is the cross section for
photo-disintegration of a nucleus with baryon number $A$ by a photon of energy
$\varepsilon'$ in the rest frame of the nucleus. We have found that
for the considerations in the present work, the GDR can be
safely approximated by the single pole of the narrow-width approximation,
\begin{equation}
\sigma_A(\varepsilon') = \pi\,\,\sigma_0\,\,  \frac{\Gamma}{2} \,\,
\delta(\varepsilon' - \varepsilon_0)\, ,
\label{sigma}
\end{equation}
where $\sigma_0 =
1.45\times 10^{-27} \, A~{\rm cm}^2$, $\Gamma = 8
\times 10^6~{\rm eV}$, and $\varepsilon_0 = 42.65 \times 10^6 A^{-0.21} \, (9.25
\times 10^5 A^{2.433})~{\rm
  eV},$ for $A > 4$ ($A\leq 4$)~\cite{Karakula:1993he}. The factor of
1/2 is introduced to match the integral (i.e. total cross section) of
the Breit-Wigner and the delta function. Inserting  (\ref{sigma}) into (\ref{rate}) we obtain
\begin{equation}
\frac{1}{\tau_{\rm int}}  \approx 
\frac{c\, \pi \, \sigma_0 \,\varepsilon_0\, \Gamma}{4 \gamma^2}
\int_{\varepsilon_0/2 \gamma}^\infty \frac{d\varepsilon}{\varepsilon^2}\,\,
  \frac{d n(\varepsilon)}{d\varepsilon} \,.
\label{kk}
\end{equation}
For the CMB,
\begin{equation}
\frac{d n(\varepsilon)}{d\varepsilon} =   \frac{1}{(\hslash
c)^3} \ \left(\frac{\varepsilon}{\pi} \right)^2 \
\left[e^{\varepsilon/T}-1 \right]^{-1} \,\,,
\label{nBE}
\end{equation}
and so (\ref{kk}) becomes~\cite{Anchordoqui:2006pd}
\begin{equation}
\frac{1}{\tau_{\rm int}^{\rm CMB}}  \approx   \frac{1}{\hslash^3 c^2} \ \frac{\sigma_0 \,\varepsilon_0\,
\Gamma\,T}{4 \gamma^2 \pi} \,\,
\left| \ln \left(1 - e^{-\varepsilon_0/2 \gamma T}\right) \right| \,, 
\label{RBE}
\end{equation}
with  $T = 2.3 \times 10^{-4}~{\rm eV}$. Following~\cite{Epele:1998ia}, we
parametrize the spectral density of the cosmic IR background as,
\begin{equation}
\frac{dn(\varepsilon)}{d \varepsilon}  \simeq 1.1 \times 10^{-4} \
\left(\frac{\varepsilon}{{\rm eV}} \right)^{-2.5}~{\rm cm^{-3} \ eV^{-1}} \,,
\end{equation}
where $2 \times 10^{-3} < \varepsilon/{\rm eV} < 0.8$. For $\gamma
\agt  10^9$, the interaction
time of a nucleus scattering off the IR  is found to be
\begin{equation}
\frac{1}{\tau_{\rm int}^{\rm IR}} \approx  8 \times 10^{-6}  \,
\left(\frac{\sigma_0}{{\rm cm}^2} \right)
  \, \left(\frac{\Gamma}{{\rm eV}}\right) \ \left(\frac{\epsilon_0}{{\rm
        eV}}\right)^{-2.5} \, \gamma^{1.5}~{\rm s}^{-1} \, . 
\label{arielH}
\end{equation}
Using numerical integration we have verified that, in the energy range
of interest,  different
parametrizations of the cosmic IR background~\cite{Dominguez:2010bv,Gilmore:2011ks} modify the interaction time
scale given in (\ref{arielH}) by $\alt 20\%$.

\begin{figure}[t]
    \postscript{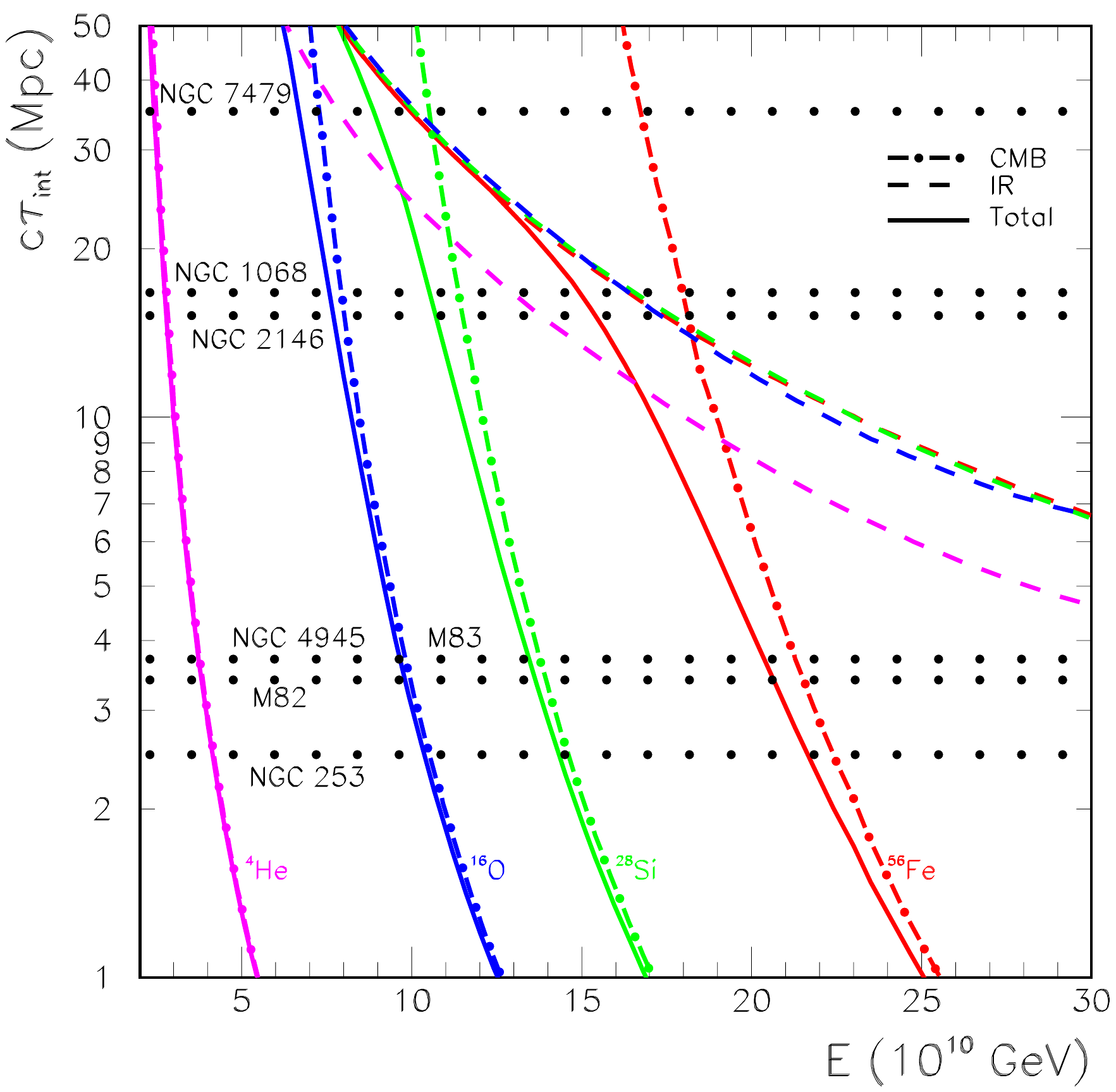}{0.99}
\caption{Photo-disintegration mean free path on the CMB and IR photon
  fields for various nuclear species. The horizontal dotted lines
  indicate the distance to nearby 
starburst galaxies in the \textit{Fermi}-LAT
catalog~\cite{Ackermann:2012vca}, with flux emission (or upper limit)
in the gamma ray band $0.1 < E/{\rm GeV} < 100$ bigger than $5 \times
10^{-9}~{\rm cm^{-2} \,  s^{-1}}$.  Starburst galaxies provide the
example of UHECR emitters which we employ.
\label{fig:4}}
\end{figure}

Our conclusions are encapsulated in Fig.~\ref{fig:4}, where we show
the mean free path of UHECR nuclei scattering off the pervasive cosmic
radiation fields.  One sees that the interaction mean free path (mfp)
decreases rapidly with increasing energy, and increases rapidly with
increasing nuclear composition:
\begin{itemize}
\item at $E = 10^{10.6}$~GeV, the mfp for ionized helium ($^4$He) is
  about 3~Mpc, while at $10^{10.85}~{\rm GeV}$ it is nil; 
\item at  $E = 10^{11}~{\rm GeV}$, the mfp for ionized oxygen ($^{16}$O) is about 4~Mpc, while at $10^{11.2}$~GeV it is nil; 
\item It $E =10^{11.2}$~GeV, the mfp for ionized silicon ($^{28}$Si)
  is about 2.5~Mpc, while at $10^{11.3}~{\rm GeV}$ it is nil;
\item etcetera, until finally we reach ionized iron ($^{56}$Fe) where
  the mfp at $E = 10^{11.3}$~GeV is about 4~Mpc, 
while at $10^{10.44}~{\rm GeV}$ it too is nil.
\end{itemize}
Thus, we have a {\it cosmic mass spectrometer.}
From sources at increasing distance, fewer and heavier nuclei at highest energies are expected to reach Earth.
The main features in the energy evolution of the
abundance of various nuclear species on Earth
can be summarized as follows: 
\begin{itemize}
\item the contribution of $^4$He should decrease with rising energy and then essentially disappear above
about $10^{10.7}~{\rm GeV}$; 
\item on average, only species heavier than $^{16}$O can contribute to the observed flux on 
Earth above $10^{11}~{\rm GeV}$, with nuclear species lighter than $^{28}$Si
highly suppressed at $10^{11.6}~{\rm GeV}$; 
\item the mean flux of iron nuclei becomes suppressed somewhat below $10^{11.4}~{\rm GeV}$. 
This is the maximum average energy expected on Earth, and is in agreement at
the $1\sigma$ level with Fly's Eye observations~~\cite{Bird:1994mp}. 
\end{itemize}
The three considerations enumerated above are similar to those
obtained assuming a continuous source distribution, with cutoff at
about 3~Mpc~\cite{Allard:2008gj,Allard:2011aa,Wykes:2017nno}.  However, the GZK
energy losses follow a trend with increasing energy similar to that
expected for acceleration capability of the sources, which grows
linearly in the charge $Ze$ of the nucleus. Namely, as the sources
start to run out of power, the contribution to the emission spectrum
of light and intermediate mass nuclei should decrease as the energy
increases. Thus, additional information on the abundance of the
different nuclear species on Earth from other nearby sources would be
needed in order to distinguish between the two scenarios. Such
additional information is provided by considering the discreteness of
UHECR emitters. 

The fractional energy loss per collision decreases with
increasing baryon number, and so we expect large fluctuations over the
mean at the highest energies. Moreover, one should expect additional
background from residual nuclear fragments created by the
propagation of heavy nuclei from very distant sources. However, our
concluding remarks could be verified by analyzing the distribution of
arrival directions, because {\it for a given CR with
  charged $Ze$, the higher the energy the smaller the bending, and therefore, 
the smaller the angle between the incident CR and the true source}.

Strictly speaking, for any potential cluster of trans-GZK events
appearing in the distribution of arrival directions we can use the
information on the nuclear composition of each independent event to
determine the distance to the source of the cluster through a
multi-parameter likelihood analysis. The set of free source parameters
involved in the data analysis, containing all the relevant guidelines
to vary the incident flux and information on the nuclear composition,
are: {\it (i)}~the flux normalization, {\it (ii)}~the spectral index
of the power-law fit, {\it (iii)}~the maximum energy, {\it (iv)}~the
admixture of nuclear species, and {\it (v)}~the distance (more
precisely, the time of flight). The fit is constrained by propagation
effects. The critical task of deriving a unique analytic relation for
the likelihood function ${\cal L}$ between the observed and emitted
nuclear species becomes workable, because energy losses are entirely
dominated by photo-disintegration. By the maximization of ${\cal L}$
in terms of the free parameters we can estimate the most likely value
for those parameters. The most likely value of the source distance can
then be combined with the distribution of arrival direction in the
cluster to search for possible correlation with source catalogs in
3-dimensions.  Discrimination between the two classes of models can be
done on an statistical basis by comparing the best parameter values,
particularly the maximum energy at the source of each individual
cluster.

Before proceeding, we pause to note an essential
difference between the method proposed herein and the traditional
technique used to search for the degree of correlation between the
arrival directions of UHECRs and source catalogs. On the one hand, in
standard cross-correlation analyses one has to impose a selection
criteria on potential sources to select a subsample of the catalog
which corresponds to a given class of objects. Moreover, usually one
has to impose very selective cuts on source parameters to select the
most powerful objects in the catalog. On the other hand, in the
analysis proposed herein we first conduct a search for the distance to the
origin of a given potential cluster through the multi-parameter likelihood analysis,
and so all source parameters are fit to the data. One can then search for
cross-correlations with catalogs in 3-dimensions and so there is no
need for selection of a particular class of sources. Our method will
automatically search for all sources of UHECRs independently of their
various possible {\it origins}.

We now turn to discuss the sensitivity of this analysis for existing
and future UHECR observatories. Determination of the source distance
with extremely high precision requires UHECR measurements with a large
exposure and first-rate $X_{\rm max}$ resolution (defined as the rms
of the distribution $X_{\rm max}^{\rm reconstructed} - X_{\rm
  max}^{\rm true}$). For example, discrimination between light and
heavy nuclei requires an $X_{\rm max}$ resolution of about $50~{\rm
  g/cm}^2$, whereas the discrimination between medium mass and heavy
nuclei requires about $20~{\rm g/cm}^2$~\cite{Aab:2016zth}.  For
AugerPrime, the $X_{\rm max}$ resolution ranges from $15~{\rm g/cm}^2$
for fluorescence detection to about $50~{\rm g/cm}^2$ for measurements
of the surface detector array~\cite{Aab:2016vlz}.  Note that because
of the $\approx 15\%$ duty cycle of fluorescence facilities only a
subsample of Auger events would have $X_{\rm max}$ measurements with
$15~{\rm g/cm}^2$ resolution. When information on $X_{\rm max}$ is
combined with measurements of $N_\mu$ and the muon shower maximum
$X_{\rm max}^\mu$, the resolution of the Auger surface detector can be
significantly improved to about $40~{\rm g/cm}^2$ at $10^{10}~{\rm
  GeV}$, reaching $\sim 25~{\rm g/cm}^2$ at $10^{11}~{\rm
  GeV}$)~\cite{Aab:2016vlz}.  The Probe Of Extreme Multi-Messenger
Astrophysics (POEMMA) is being designed to achieve orders-of-magnitude
increase in statistics of observed UHECRs beyond $10^{10.6}~{\rm
  GeV}$~\cite{Olinto:2017xbi}. POEMMA stereo observations will have a
large enough sample of well reconstructed UHECR events, with a
resolution of at least $\sim 60~{\rm g/cm}^2$. This will be enough to
distinguish protons from heavy nuclei. At this point we should note
that results of air shower simulations show that shower-to-shower
fluctuations in $X_{\rm max}$ are large, and even extreme compositions
like pure proton and pure iron have a considerable overlap in their
$X_{\rm max}$-distributions. This introduces a theoretical systematic
uncertainty in the proposed likelihood analysis.  We look forward to the
nuclear composition being presented for each event~\cite{Aab:2017cgk}, so that the
analysis proposed in this paper can be undertaken.

\section{\textbf{\textit{Cepa stratis}}}
\label{cepastratis}

Hitherto we have adopted a pragmatic approach and avoided the details
of theoretical modeling of magnetic deflections. Magnetic fields are
not well constrained by current data, but if we adopt recent models of
the Galactic magnetic
field~\cite{Pshirkov:2011um,Jansson:2012pc,Jansson:2012rt,Unger:2017kfh},
typical values of the deflections of UHECRs crossing the Galaxy are
$\sim 10^\circ$ for $E/Z = 10^{10}~{\rm GeV}$, depending on the
direction considered~\cite{Farrar:2014hma,Aab:2017tyv,Farrar:2017lhm}.\footnote{Extragalactic
  magnetic fields may also be relevant for UHECR propagation in the
  intergalactic space; see e.g.,~\cite{vanVliet:2017oci}.}

When the average magnetic field deflection is combined with the energy
losses shown in Fig.~\ref{fig:4}, we can conclude that:
\begin{itemize}
\item  Granting that the extragalactic UHECR population seem to include a significant
  fraction of nuclei we still expect to observe an anisotropy, due to
  the anisotropic distribution of matter within the GZK horizon.
\item In particular, for $10^{10.6} \alt E/{\rm GeV} \alt 10^{11}$, we
  expect to observed excesses in $\sim 15^\circ$ regions of
  the sky centered at nearby sources, associated with nuclei of $Z
  \alt 10$.
\item For $E \agt 10^{11}~{\rm GeV}$, the population of UHECR nuclei
  observed at Earth must consist mostly of baryons with $Z \agt 10$,
  or else protons. This statement can be verified by inspection of
  Fig.~\ref{fig:4}. Nuclei heavier than neon would suffer too much of
  a deflection to be contained in a $\sim 15^\circ$ hot spot, and in
  some cases (e.g. iron nuclei) may completely camouflage the exact
  location of the sources.
\item The {\it cepa stratis} structure inherent to the search of
  nucleus-emitting-sources is in sharp contrast to the quest for UHECR
  proton accelerators, in which the higher the energy of the proton
  the smaller the bending on the magnetic field and therefore the
  smaller the angle between the incident proton direction at Earth and
  the line-of-sight to the true source. The future POEMMA mission,
  which is expected to monitor the full sky with an extremely-fast,
  highly-pixelized, immensely-large aperture, will become the
  optimal instrument to identify the various types of UHECR sources.
\item The potential observation of hot spots from nearby sources with
  onion layers that increase in size with rising energy,
  or else the observation of compact hot spots that become denser, compressed, and
  increase the significance level at
  $E \agt 10^{11}~{\rm GeV}$ could provide a determination of the
  UHECR nuclear composition. Such UHECR species determination would be
  completely independent of CR  air shower properties, and consequently not
  affected by the large systematic uncertainties introduced by models
  of hadronic interactions at ultrahigh energies.
\end{itemize}

In closing, we stress that the {\it cepa stratis} structure originates
in the peculiar balance of magnetic field deflections and energy
losses of UHECR nuclei.  As a consequence, this is a global
effect. Whether a particular source of UHECR nuclei could be exposed
through its onion layers would depend on the direction considered in
the sky.

All source types are represented within the GZK horizon for heavy
nuclei, so all source types are {\sl a priori} candidates for the
nearby exploration.  However, starburst galaxies are perhaps the
leading candidate for UHECR nucleus-emitting sources, and so we will
take them as the example for our study.  Readers not so interested in
the details of the starburst galaxies example which we develop next,
can skip to the summary section.

\section{
Source Example: Starburst Galaxies}
\label{starburst}

It has long been suspected that galaxies with bursts of massive star
formation (starbursts) have the power to accelerate UHECR
nuclei~\cite{Anchordoqui:1999cu,Torres:2004hk}. It has also been noted that the
arrival directions of the highest energy cosmic rays recorded by the
Fly's Eye~\cite{Bird:1994mp} ,
AGASA~\cite{Hayashida:1994hb,Takeda:1998ps}, and
Yakutsk~\cite{Knurenko:2017zqd} experiments can be traced back to the
two nearest starbursts: M82 and NGC
253~\cite{Anchordoqui:2002dj}. 

\begin{figure}[tbp]
    \postscript{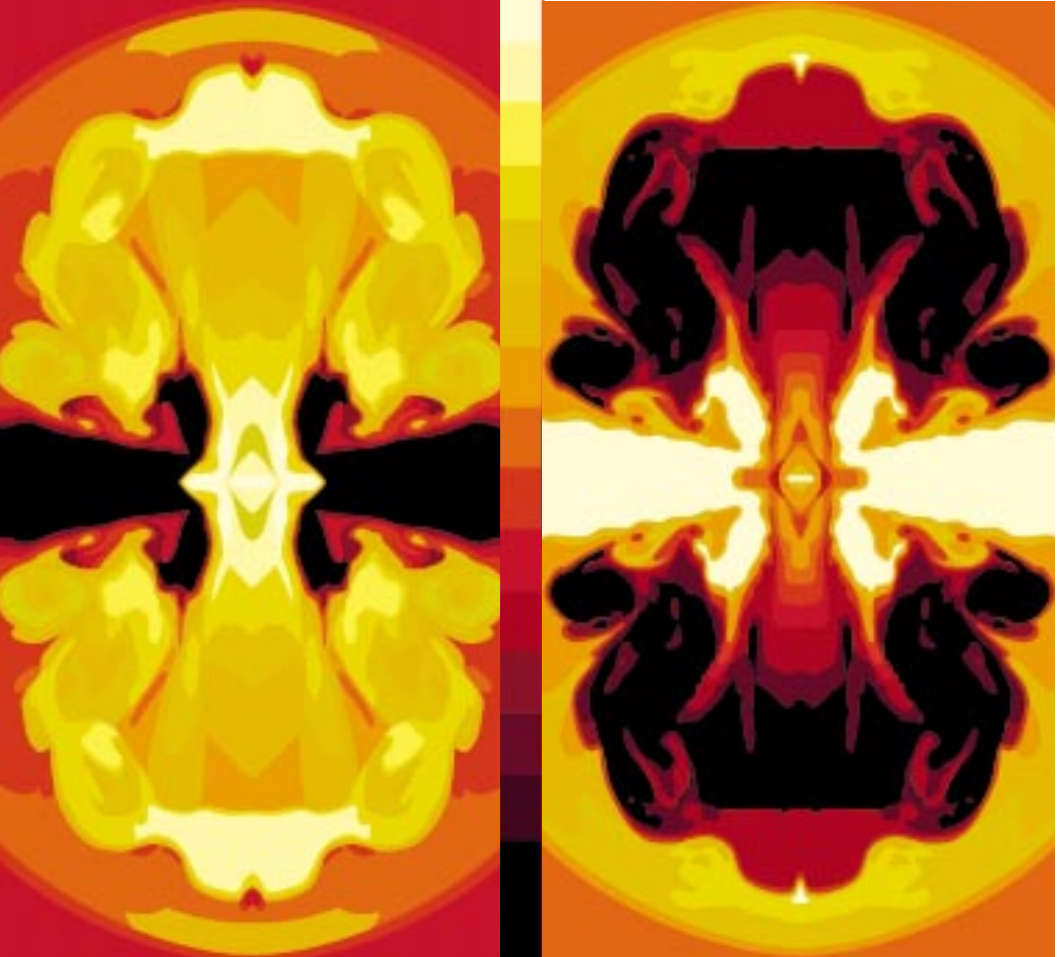}{0.99}
    \caption{Numerical simulation of a starburst's superwind. {\it
        Left panel.} Temperature map (bright = hot) showing how the
      hot gas emanating from the nucleus displaces the cooler galactic
      gas around it.  {\it Right panel.} Gas density map (bright =
      dense) showing the inhomogeneous outflow along the rotation axis
      of the disk composed of a series of
      hot, dense, and fast shock fronts of material that are trailed
      by gas which has expanded, cooled, and slowed down.  This figure
      is courtesy of Gerald Cecil~\cite{Veilleux}.
\label{fig:1}}
\end{figure}

\begin{figure}[tbp]
    \postscript{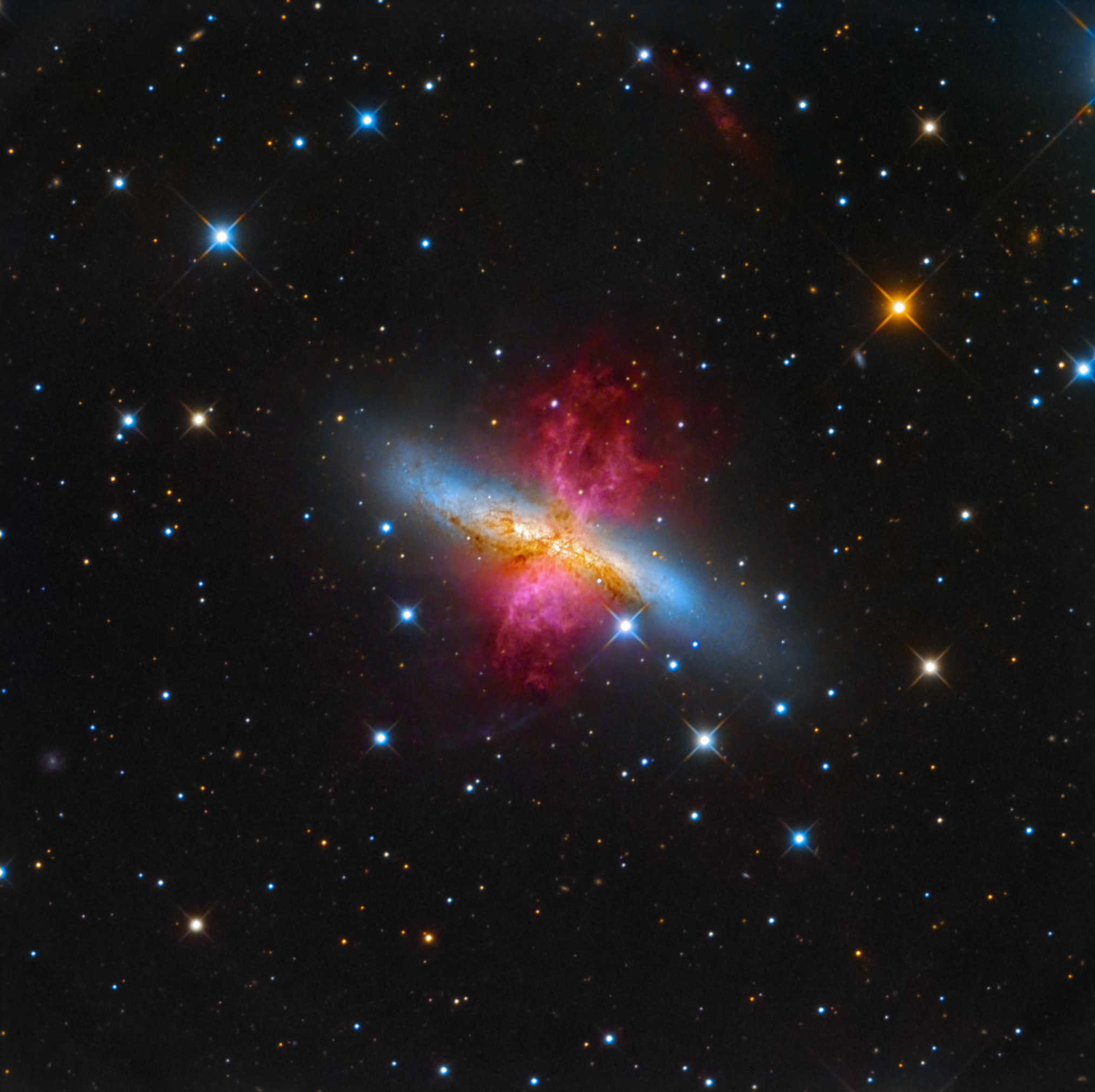}{0.99}
    \caption{Telescopic snapshot of M82. Shortly after the
      identification of optical emission-line filaments~\cite{Lynds}
      it became widely accepted that M82 is the archetype starburst
      galaxy~\cite{OConnell}.  The huge lanes of dust that crisscross
      the disk of M82 are the telltale signature of the flurry of star
      formation. Winds from massive stars and blasts from supernova
      explosions have created a strong superwind of galactic-scale,
      which is spewing knotty filaments of hydrogen and nitrogen
      gas~\cite{Shopbell:1997dj,Suchkov}.  The red-glowing outwardly
      expanding filaments featuring the H$\alpha$ emission provide
      direct evidence for the galactic-scale superwind emanating from
      the central region to the outer halo area. This figure is
      courtesy of Leonardo Orazi.
  \label{fig:2}}
\end{figure}

Starburst galaxies feature strong infrared emission by dust associated
with high levels of interstellar extinction, strong UV spectra from
the Lyman-$\alpha$ emission of hot OB stars, and considerable radio
emission produced by recent supernova remnants. The luminosity of the
Balmer lines, primarily H$\alpha$ and H$\beta$, gives a measure of the
star formation rate.  The central regions of starburst galaxies can be
orders of magnitude brighter than those of normal spiral galaxies.
From such an active region, it is known~\cite{Veilleux:2005ia,Heckman} that a
galactic-scale superwind is driven by the collective effect of
supernovae and winds from massive stars.  The high supernovae rate
creates a cavity of hot gas ($\sim 10^8~{\rm K}$) whose cooling time
is much greater than the expansion time
scale~\cite{Rieke:1980xt,Chevalier:1985pc,Heckman:1990fe,Hoopes:2003sh}. Since the wind is
sufficiently powerful, it can blow out of the interstellar medium of
the galaxy as a hot bubble.  As the cavity expands a strong shock
front is formed on the contact surface with the cool interstellar
medium.  Shock interactions with low and high density clouds produce
X-ray continuum and optical line emission, respectively.  This model
is supported by numerical simulations, an example of which is shown in
Fig.~\ref{fig:1}.  The majority of nearby superwinds have been
discovered using optical imaging and spectroscopy to identify
galaxy-sized outflows with velocities in excess of several hundred to
a few thousand kilometers per second, and in most cases (e.g. M82
shown in Fig.~\ref{fig:2}) directly imaging structures aligned with
the host galaxy's transverse axis.

Because of the high prevalence of supernovae, starbursts should
possess a large density of newly-born pulsars. Due to their important
rotational and magnetic energy reservoirs these young neutron stars,
with their metal-rich surfaces, have been explored as a potential
engine for UHECR acceleration~\cite{Blasi:2000xm,Fang:2012rx}. The
acceleration mechanism in a young neutron star is unipolar induction:
In the out-flowing relativistic plasma, the combination of the fast
star rotation and its strong magnetic field can induce, in principle,
potential differences of order $\Delta V = \Omega^2 \mu/c^2$, where
$\mu = B_\star R^3_*/2$ is the magnetic dipole moment, $B$ is the surface
dipole field strength, $\Omega$ is the rotation frequency, and $R_*
\sim 10~{\rm km}$ the star radius. The fastest spinning young neutron
stars exhibit large magnetic fields typically in the range $10^{12} \lesssim
B_\star/{\rm G} \lesssim 10^{13}$, yielding $\mu \sim 10^{30.5}~{\rm cgs}$.
Provided that charged particles can experience a fraction
$\eta$ of that potential, they will be accelerated to the
energy~\cite{Blasi:2000xm}
\begin{eqnarray}
E (\Omega) & = & Ze \, \eta \, \Delta V  \nonumber \\ 
& \sim &  10^{11} \,
   \frac{Z}{26} \ \frac{\eta}{0.03} \ \left(\frac{\Omega}{10^4~{\rm s^{-1}}}\right)^2 \frac{\mu}{10^{30.5}~{\rm cgs}}~{\rm GeV}\,.
\label{ocho}
\end{eqnarray}
The fiducial value of $\Omega$ adopted in (\ref{ocho}) corresponds to
the exceptionally fast spinning young pulsars. The majority of pulsars
are born spinning slower. Indeed, the distribution of pulsar-birth
spin periods, $f(P=2 \pi/\Omega)$, is Gaussian, centered at 300~ms, with
standard deviation of 150~ms~\cite{FaucherGiguere:2005ny}.  Note that
most of the pulsars would accelerate heavy nuclei up a few
$10^{6}~{\rm GeV}$. Proto-pulsars spinning initially with $P \approx
6~{\rm ms}$~\cite{Haensel:1999mi} could reach $E \sim 10^{9}~{\rm
  GeV}$, which is roughly the maximum energy of Galactic cosmic
rays~\cite{Fang:2013cba}.

Neutron-star surfaces are thought to be composed of anisotropic,
tightly-bound condensed matter. The crust of neutron stars extends
down to about 1~km below the surface, with densities ranging from a
few ${\rm g/cm^3}$ on the exterior surface up to nuclear density
$10^{14}~{\rm g/cm^3}$ in the interior~\cite{Chamel:2008ca}.  The
outermost layers of a neutron stars are composed of iron. At densities
$\agt 10^4~{\rm g/cm^3}$, the atoms are fully ionized due to the
pressure of the upper layers. The free electrons are degenerate and
become relativistic at densities $> 10^6~{\rm g/cm^3}$. With
increasing densities, the nuclei are more and more neutron rich owing
to electron captures which convert protons into neutrons. This
neutronization of the matter leads to the existence of a neutron ocean
which permeates the inner layers of the crust at densities $\agt
10^{11}~{\rm g/cm^3}$. The crust dissolves into a uniform mixture of
neutron, protons and electrons when the density reaches about
$10^{14}~{\rm g/cm^3}$. $^{56}$Fe ions can thus be stripped off the
surface and be accelerated through unipolar induction.  A recent
study~\cite{Kotera:2015pya} demonstrates that for the most reasonable
range of neutron star surface temperatures ($T < 10^7~{\rm K}$), a
large fraction of nuclei survive {\it complete} photo-disintegration in the
hostile environment sustained by the thermal radiation field from the
star. However, the  apparently inconsequential photo-disintegration
losses could still be enough to produce a mixed nuclear
composition at the source, with a non-negligible CNO component.  The spectrum of
accelerated UHECRs is determined by the evolution of the rotational
frequency: As the star spins down, the energy of the ejected cosmic
ray particles decreases. As a consequence, the total fluence of UHECRs
accelerated in the neutron star magnetosphere is very hard, with
spectrum $\propto E^{-1}$. Interestingly, as we noted in
Sec.~\ref{intro}, simultaneously reproducing Auger data on the
spectrum together with the observed nuclear composition requires a
hard source spectrum~\cite{Aloisio:2013hya,Unger:2015laa,Aab:2016zth}.

After the nuclei escape from the central region of the galaxy, with
$10^6 \alt E/{\rm GeV} \alt 10^{9}$, they are injected into the
galactic-scale superwind and could potentially experience diffusive shock
acceleration~\cite{Bell:1978zc,Bell:1978fj,Lagage:1983zz,Drury:1983zz,Blandford:1987pw}. Diffusive shock acceleration is a first-order Fermi acceleration
process~\cite{Fermi:1949ee} in which charged particles increase their
energy by crossing the shock front multiple times, scattering off
turbulence in the magnetic field $B$. The maximum achievable energy is
obtained by setting the acceleration and flow time scales equal to
each other~\cite{Gaisser:1990vg}. The constraint due to the finite lifetime of the shock
yields,
\begin{equation}
E_{\rm max} \sim \frac{1}{12} \ Ze \  B \ v_{\rm sw}^2 \
\tau \, ,
\label{Emax}
\end{equation} 
where $v_{\rm sw} \sim \sqrt{2 \dot E_{\rm sw}/ \dot M_{\rm sw}}$ is
the asymptotic speed of the outflow in the superwind, $\dot E_{\rm
  sw}$ and $\dot M_{\rm sw}$ are respectively the energy and mass
injection rates inside the spherical volume of the starburst region,
and $\tau$ is the lifetime of the starburst, and where we have assumed
a strong shock with a shock compression ratio $r = 8$ (associated to
the adiabatic index of a polyatomic gas)~\cite{windy}. In (\ref{Emax})
it was implicitly assumed that the magnetic field is parallel to the
shock normal. Injecting additional constraints into the model may
reduce the maximum achievable
energy~\cite{Zweibel:2002ta,Bustard:2016swa}.

To get some idea of the orders of magnitude involved in (\ref{Emax}),
a very rough estimate can be made assuming M82 typifies the population
of nearby starburst galaxies. The predicted kinetic energy and mass injection
rates, derived from the measured IR luminosity, are $3
\times 10^{42}~{\rm erg \ s^{-1}}$ and $3 M_\odot~{\rm yr}^{-1}$,
respectively~\cite{Heckman}.  The gamma-ray, radio, and far infrared
spectra of nearby starbursts seem to favor a synchrotron cooling
timescale for electrons that is much shorter than their escape
time~\cite{Paglione:2012ma}.  It has been suggested that if electrons
cool rapidly via synchrotron radiation, the magnetic energy density of
the starburst region could be significantly higher than that expected
from equipartition arguments with comparable cosmic rays and magnetic
energy densities~\cite{Thompson:2006is}.  Indeed, if the magnetic
energy density is in rough equipartition with its hydrostatic
pressure, the implied magnetic field strength of M82 on few hundred
parsec scales would be about $1.6~{\rm mG}$. Radio continuum and
polarization observations provide an estimate of the magnetic field
strength in the halo of M82, $B \sim 35~\mu{\rm
  G}$~\cite{Krause:2014iza}. The age of the starburst phase is subject
to large uncertainties. For our calculations, we adopt $\tau \sim
350~{\rm My}$, which is in the lower end of the average
ages~\cite{McQuinn:2010kn}.  All in all, substituting these figures
into (\ref{Emax}) we obtain
\begin{equation}
E_{\rm max}
\sim Z \,  10^{10}~{\rm GeV} \, .
\label{Emax2}
\end{equation}
Note that (\ref{Emax2}) is consistent with the Hillas criterion~\cite{Hillas:1985is}, as the maximum
energy of confined nuclei is found
to be
\begin{equation}
E_{\rm max} \simeq 10^9 \ Z \ \frac{B}{\mu G} \ \frac{R_{\rm sh}}{{\rm
    kpc}}~{\rm GeV} \ ,
\end{equation}
where
\begin{equation}
R_{\rm sh} \sim \sqrt{\frac{\dot M_{\rm sw} \ v_{\rm sw} }{2 \Omega \
    P_{\rm halo}}} \sim 8~{\rm kpc} 
\label{shradius}
\end{equation}
is the shock radius,  $\Omega$ is the solid angle subtended by the outflow
cones, and $P_{\rm
  halo}$ is  the pressure
inside the halo~\cite{Lacki:2013zsa}. In the estimate of (\ref{shradius}) we have
  taken $\Omega \sim \pi$ and $P_{\rm halo} \sim 10^{-14}~{\rm erg} \, {\rm
  cm}^{-3}$~\cite{Shukurov:2002hh}. The source emission spectrum would remain hard provided its shape is driven by
UHECR nucleus leakage from the boundaries of the shock (a.k.a direct
escape)~\cite{Baerwald:2013pu}. Note that for $r=8$, we expect a hard
\mbox{spectrum ($\propto E^{-1.4}$)} at the sources~\cite{windy}.

The first generation of UHECR observatories reported several events
above $10^{11.3}~{\rm GeV}$, with no indication of a GZK cutoff. This
would imply that if the UHECR were heavy nuclei, these sources must be
nearby, less than about 3~Mpc away from Earth.  More recent data,
however, show a clear cutoff around $10^{10.6}~{\rm GeV}$.  These
newer results then imply that UHECR nuclei could originate in more
distant sources, as far away as about 50~Mpc, the canonical GZK
horizon.  The anisotropy study in~\cite{Anchordoqui:2002dj} was based
on data from first-generation observatories; newer data relax both the
nuclear composition and the distance to sources.

\begin{figure}[tbp]
    \postscript{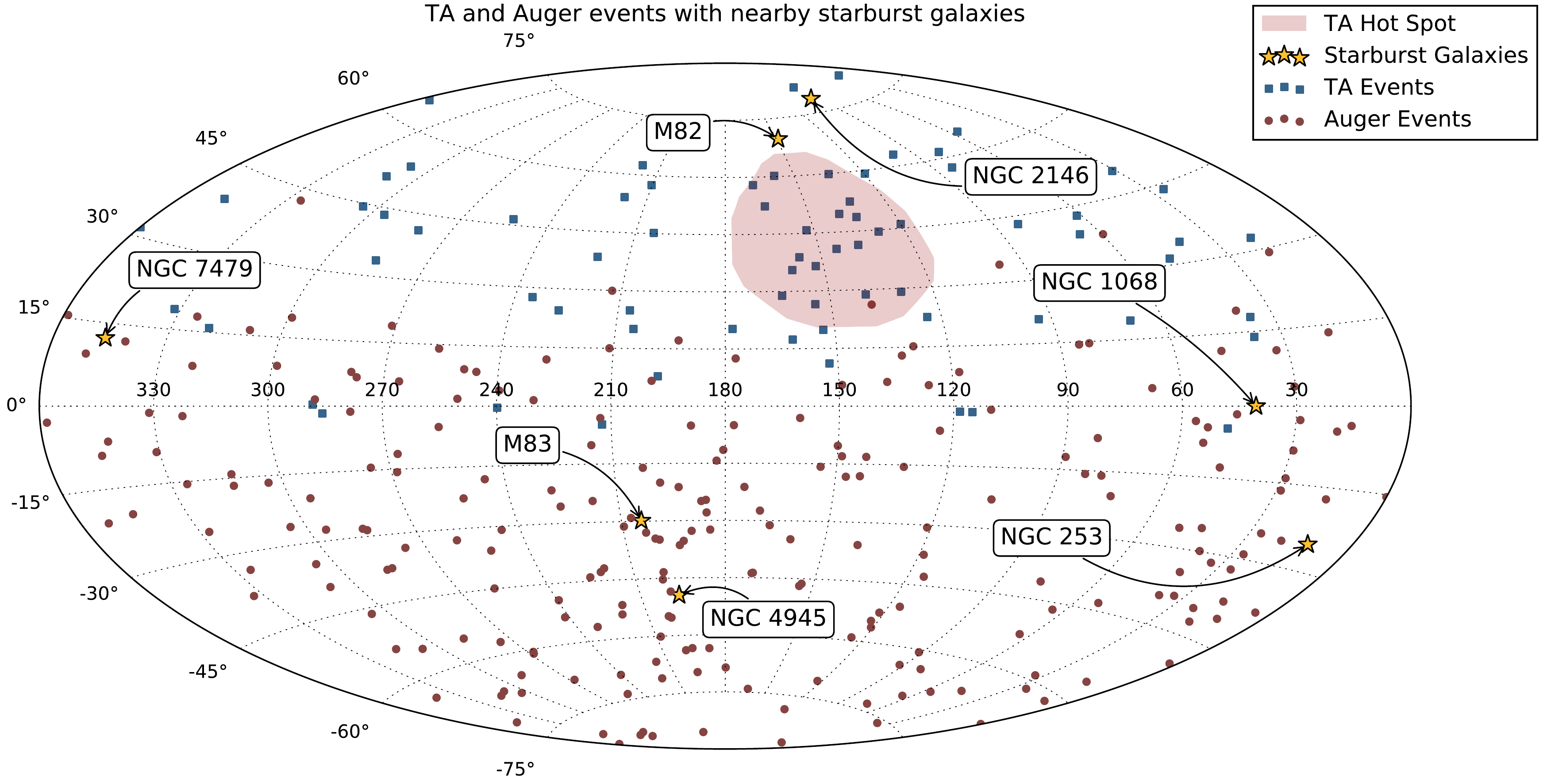}{0.99}
\caption{Comparison of UHECR event locations with nearby starburst
  galaxies in equatorial coordinates, with R.A. 
increasing from right to left. The circles indicate the arrival directions of 231 events with $E >
52~{\rm EeV}$ and zenith angle $\theta < 80^\circ$ detected by the
Pierre Auger Observatory from 2004 January 1 up to 2014 March
31~\cite{PierreAuger:2014yba}. The squares indicate the arrival
directions of 72 events with $E > 57~{\rm EeV}$ and $\theta <
55^\circ$ recorded from 2008 May 11 to 2013 May 4 with TA~\cite{Abbasi:2014lda}. The
stars indicate the location of nearby (distance $< 50~{\rm Mpc}$)
starburst galaxies. The shaded region delimits the TA hot-spot.
\label{fig:3}}
\end{figure}

An apparent correlation between UHECRs and nearby starbursts is
visible in~Fig.~\ref{fig:3}.  As a matter of fact, the Pierre Auger
Collaboration has recently reported an indication of a possible
correlation between UHECRs ($E > 3.9 \times 10^{10}~{\rm GeV}$) and
starburst galaxies, with an {\it a posteriori} chance probability in
an isotropic CR sky of $p_{\rm Auger} = 4.2 \times 10^{-5}$,
corresponding to a 1-sided Gaussian significance of $4
\sigma$~\cite{Aab:2017njo,ObservatoryMichaelUngerforthePierreAuger:2017fhr}.\footnote{Note that the significance of
  this {\it a posteriori} study does not account for the previous
  searches made within the Auger Collaboration and those made by
  others~\cite{Aab:2017njo,ObservatoryMichaelUngerforthePierreAuger:2017fhr}.} In addition, the possible association
of the TA hot spot with M82 has not gone
unnoticed~\cite{Anchordoqui:2014yva,He:2014mqa,Pfeffer:2015idq}.  The
multiplicative $p$-value for the two non-correlated observations is
\begin{equation}
p =
p_{\rm TA} \otimes p_{\rm Auger} = 1.5 \times 10^{-8}, 
\label{p-value}
\end{equation}
yielding a statistical significance $\agt 5 \sigma$.  However, caution
must be exercised in all-sky comparisons~\cite{Globus:2016gvy}.
Moreover, in (\ref{p-value}) we have combined a catalog-based
cross-correlated search (Auger) with a blind search (TA). Therefore,
(\ref{p-value}) provides a rough estimate of the statistical
significance under the strong assumption that M82 (which is at the
border of the excess of TA events) is the only source contributing to
the TA hot spot. It is clear that new data are needed to confirm the
suggested correlation.  It is also clear that existing data concerning
the distribution of arrival directions {\it do not rule out} the
possibility of starbursts as UHECR emitters. Moreover, the hard
emission spectrum from starburst galaxies renders them plausible
candidate sources. Altogether, starbursts galaxies (shown in
Fig.~\ref{fig:4}) emerge as the prominent example of source type.

We now turn to discuss how the joint study of anisotropy signals and
nuclear composition could offer valuable clues for shedding light on
nucleus-emitting-sources, such as starburst galaxies.  A particularly
interesting aspect of this study, which complements the likelihood fit
proposed in Sec.~\ref{losses}, is to analyze the pattern of
anisotropies as a function of energy~\cite{Lemoine:2009pw}. Note that
if a source produces an anisotropy signal at energy $E$ with cosmic
ray nuclei of charge $Ze$, it should also produce a similar anisotropy
pattern at energies $E/Z$ via the proton component that is emitted
along with the nuclei, given that the trajectory of cosmic rays within
a magnetic field is only rigidity-dependent. Note that the central
engine of the acceleration model discussed above is the neutron star
surface, where iron nuclei are stripped off and can be accelerated to
ultrahigh energies in a two-step process, without suffering
catastrophic interactions. This means that the accompanying proton
flux would be largely negligible. As noted in~\cite{Liu:2013ppa},
secondary protons produced during propagation could also create an
anisotropy pattern in the ``low'' energy regime. This sets a constraint on
the maximum distance to nucleus-emitting-sources.  Making the extreme
assumption that the source does not emit any proton, the source(s)
responsible for TA and Auger anisotropies should lie closer than $\sim
20$ to 30, 80 to 100, and 180 to 200~Mpc, if the anisotropy signal is
mainly composed of oxygen, silicon and iron nuclei, respectively~\cite{Liu:2013ppa}. We
note that the starburst galaxies dominating the anisotropy signal are
all at a distance $\alt 20~{\rm Mpc}$, and consequently the model would
also automatically satisfy this forceful constraint.

\begin{table}
\caption{Nearby starbursts, their location in the sky (SL), the CR
  energy ($E/(10^{10}~{\rm GeV})$), the CR arrival direction (CR-AD),
  and the angular distance ($\delta$) between SL and CR-AD for
   UHECRs which are within $\alt 15^\circ$ of the starbursts. If a given CR has
  two starbursts within $15^\circ$ only the nearest source is
  given. SL and CR-AD are in equatorial
  coordinates (in degrees); $\delta$ is also given in degrees. \label{table}}
\begin{tabular}{c c c c c }
\hline \hline
Starburst  & SL & $E/(10^{10}~{\rm GeV})$ & CR-AD & $\delta$ 
\\
\hline
NGC 2146~~&~~$(94.7, 78.4)$~~&~~$\phantom{1}6.54$~~&~~$(87.6,81.5)$~~&~~$\phantom{1}3.4$~~ \\        
NGC 2146 & $(94.7, 78.4)$ &  $\phantom{1}6.42$ &  $(22.5, 80.1)$ &   12.7 \\ 
NGC 1068 & $(40.7,0.0)$ & $\phantom{1}6.05$ & $(47.7,-4.7)$ &  $\phantom{1}8.2$ \\
NGC 1068 & $(40.7,0.0)$ & $\phantom{1}6.69$   &   $(28.9,-2.7)$ &   12.1 \\
NGC 1068 & $(40.7,0.0)$ & $10.82$ &    $(45.6,-1.7)$ &   $\phantom{1}5.2$ \\
NGC 1068 & $(40.7,0.0)$ & $\phantom{1}6.77$  &  $(53.0, -4.5)$ & 13.1 \\
M82  & $(149.0, 69.7)$ & $\phantom{1}5.78$ &  $(158.6, 60.3)$ & 10.2 \\
M82  & $(149.0, 69.7)$ & $\phantom{1}7.69$ &  $(134.8, 59.8)$  & 11.5 \\   
M82  & $(149.0, 69.7)$ & $\phantom{1}8.33$ & $(168.5, 57.9)$ & 14.4 \\
NGC 7479 & $(346.2,12.3)$ & $\phantom{1}6.52$ &  $(331.65,18.85)$ &  15.5\\
NGC 7479 & $(346.2,12.3)$ & $\phantom{1}8.90$ &  $(349.9,9.3)$ &  $\phantom{1}4.7$ \\
NGC 7479 & $(346.2,12.3)$ & $\phantom{1}5.67$ &  $(358.9,15.5)$ &    12.7 \\
NGC 7479 & $(346.2,12.3)$ &$11.83$ & $(340.6,12.0)$ &    $\phantom{1}5.5$ \\ 
NGC 253 & $(11.9,-25.3)$ & $\phantom{1}6.04$  &    $(19.8,-25.5)$ &  12.4 \\
NGC 253 & $(11.9,-25.3)$ & $\phantom{1}6.47$   &   $(15.6,-17.1)$ &  12.7 \\
NGC 253 & $(11.9,-25.3)$ & $\phantom{1}5.90$   &   $(26.7,-29.1)$  & 13.6  \\
NGC 253 & $(11.9,-25.3)$ & $\phantom{1}7.37$     & $(12.3,-40.7)$  & 15.4 \\
NGC 253 & $(11.9,-25.3)$ & $\phantom{1}6.33$    &  $(26.1,-32.2)$  & 14.2 \\
NGC 253 & $(11.9,-25.3)$ & $\phantom{1}7.02$   &   $(4.6,-37.9)$   & 14.0  \\
NGC 253 & $(11.9,-25.3)$ & $\phantom{1}8.38$  &    $(26.8,-24.8)$  & 13.5 \\   
NGC 253 & $(11.9,-25.3)$ & $\phantom{1}7.12$ &     $(17.5,-37.8)$  & 13.4 \\
NGC 4945 & $(196.4,-49.5)$ & $\phantom{1}5.86$ & $(208.1,-60.1)$ &  12.6\\
NGC 4945 & $(196.4,-49.5)$ & $\phantom{1}5.21$ &  $(199.1,-48.5)$ &  $\phantom{1}2.0$ \\
NGC 4945 & $(196.4,-49.5)$ & $\phantom{1}6.95$ &  $(201.1,-55.3)$ &   $\phantom{1}6.5$ \\
NGC 4945 & $(196.4,-49.5)$ & $\phantom{1}5.95$ & $(200.9,-45.3)$ &   $\phantom{1}5.2$ \\
NGC 4945 & $(196.4,-49.5)$ & $\phantom{1}6.00$ & $(200.2,-43.4)$ &   $\phantom{1}6.6$ \\
NGC 4945 & $(196.4,-49.5)$ & $\phantom{1}6.15$  & $(219.5,-53.9)$ &   14.9 \\
NGC 4945 & $(196.4,-49.5)$ & $\phantom{1}6.19$ &  $(195.5,-63.4)$ &   13.9 \\
NGC 4945 & $(196.4,-49.5)$ & $\phantom{1}6.53$ & $(187.5,-63.5)$ &   14.8 \\
NGC 4945 & $(196.4,-49.5)$ & $\phantom{1}5.33$ &  $(202.0, -54.9)$ &  $\phantom{1}6.4$\\
NGC 4945 & $(196.4,-49.5)$ & $\phantom{1}5.86$ &  $(217.9,-51.5)$ &   13.8\\
M83 & $(204.2,-29.9)$ & $\phantom{1}5.92$ &  $(199.7, -34.9)$ & $\phantom{1}6.4$ \\   
M83 & $(204.2,-29.9)$ & $\phantom{1}7.25$ &   $(193.8,-36.4)$ & $10.9$ \\
M83 & $(204.2,-29.9)$ & $12.71$ &  $(192.8,21.2)$ & $13.4$\\
M83 & $(204.2,-29.9)$ & $\phantom{1}6.07$   &  $(192.5,-35.3)$ & $11.3$ \\
M83 & $(204.2,-29.9)$ & $\phantom{1}5.81$   & $(202.2,-16.1)$ & $13.8$ \\
M83 & $(204.2,-29.9)$ & $\phantom{1}7.40$  & $(209.6,-31.3)$  & $\phantom{1}4.8$\\
M83 & $(204.2,-29.9)$ & $\phantom{1}6.67$ & $(203.4,-33.0)$ & $\phantom{1}3.2$\\
M83 & $(204.2,-29.9)$ & $\phantom{1}7.25$ & $(193.8,-36.4)$ & $10.9$\\
M83 & $(204.2,-29.9)$ & $\phantom{1}5.47$ & $(197.8,-20.0)$ & $11.5$\\
M83 & $(204.2,-29.9)$ & $\phantom{1}6.48$ & $(207.1,-29.1)$ & $\phantom{1}2.6$\\
M83 & $(204.2,-29.9)$ & $\phantom{1}5.57$ & $(217.1, -24.5)$ & $12.6$\\
M83 & $(204.2,-29.9)$ & $\phantom{1}62.7$ & $(200.9,-34.6)$ & $\phantom{1}5.5$\\
M83 & $(204.2,-29.9)$ & $\phantom{1}7.45$ & $(189.9,-32.7)$ & $12.6$ \\
\hline
\hline
\end{tabular}
\end{table}

In regards to the expected {\it cepa stratis} structure introduced in
Sec.~\ref{cepastratis}, we note that most UHECRs contributing to the
possible correlation with starbursts (shown in Fig.~\ref{fig:3}) have
energies below $10^{11}~{\rm GeV}$.  More specifically, in
Table~\ref{table} we display the energy and arrival directions of UHECRs which are
at $\alt 15^\circ$ degrees from our fiducial starburst galaxies. Only
3 out of the 44 events have energies in excess of $10^{11}~{\rm
  GeV}$. The highest energy event, with $E \simeq 1.3 \times
10^{11}~{\rm GeV}$, appears to correlate with M83 which is only 4~Mpc
away. The other two events are from sources which are beyond 15~Mpc,
and from which we consequently expect species heavier than silicon; 
see Fig.~\ref{fig:4}. We conclude that  there is a global
agreement between the starburst hypothesis and data, interpreting 
these two events as a natural fluctuation.

Given that a strong evidence for a correlation between the arrival directions of
UHECR and starbursts has been already established, the analysis method
discussed in Sec.~\ref{losses}  should be adapted to this particular
case. The multi-dimensional likelihood analysis to determine the
source parameters and distance  considering separate data from each
cluster is still pending. This analysis would provide complementary
information to test 
the starburst hypothesis of UHECRs. Predictions on possible mass composition can
be advanced from Fig.~\ref{fig:4}, where we show distances to the
nearest likely starburst galaxies.   We see that there are four
likely starburst candidates within 4~Mpc of Earth, and two more at
$\sim 15$~Mpc from Earth.  We get:
\begin{itemize}
\item the maximum average energy of nuclei arriving from 
NGC 2146 and/or NGC 1068 is roughly $10^{11.2}~{\rm GeV}$;
\item no ions lighter than $^{28}$Si would be observed from NGC 2146 and/or NGC 1068 
with an average energy beyond $10^{11.13}~{\rm GeV}$;
\item only nuclei lighter than neon, with a spectral cutoff $E \sim
  10^{11}$~{\rm GeV}, will appear to arrive from
the direction (i.e. within $15^\circ$ or so) of NGC 2146 and/or NGC 1068.
\end{itemize} 

Another interesting application of our method is as follows. The sources
NGC 1068 and NGC 7479 are located at about 15 and 35 Mpc,
respectively. The propagation distances then engender specific
patterns in the individual spectra and nuclear composition, e.g., no
nuclear species lighter than $^{16}$O would be expected above $E \sim
10^{10.9}~{\rm GeV}$. Moreover,  these two sources are located in a
region of the sky in which the apertures of Auger and TA overlap. As a
consequence, these sources can be used for normalization of the
starburst hypothesis. Namely, even though the CR flux in the Northern
and Southern skies may differ as a reflection of the cosmic variance,
for a given angular aperture to the line-of-sight to these sources,
the event rates should be the same for both Auger and TA. This is another
concrete example of how the proposed  multi-dimensional analysis could
provide valuable clues to unmask the origin of UHECRs.

We note that  other nearby starbursts may contribute to the UHECR
flux observed on Earth, e.g., NGC 3079 at a distance of 16.2~Mpc is
known to have a powerful large scale
superwind~\cite{Cecil:2001it,Cecil:2002xp}, and emission in the gamma
ray band $0.1 < E/{\rm GeV} < 100$ smaller than $2.2 \times
10^{-9}~{\rm cm^{-2} \, s^{-1}}$~\cite{Ackermann:2012vca}. The
presence of these additional sources would not alter our conclusions.

Finally, we note that the starburst model has implications for
ultrahigh energy neutrinos.  As shown
elsewhere~\cite{Anchordoqui:2007tn}, for very reasonable source
parameters, we expect the radiation backgrounds to hardly disintegrate
accelerated nuclei enabling a direct escape, largely without
contributing to the cosmic neutrino flux. This is consistent with the
emission of a negligible proton flux and the absence of anisotropy
patterns at energies $E/Z$, in agreement with observations.

\section{Summary}
\label{summary}
It is well known that GZK interactions of UHECR nuclei {\it en route}
to Earth favor heavy nuclei at the highest energies. The maximum
energy of the acceleration capability of sources grows linearly in $Z$
and hence also favors heavy nuclei at the highest energies.  The traditional
  bi-dimensional analyses, which simultaneously reproduce Auger data
  on the spectrum and nuclear composition, may not be capable  of distinguishing the
relative importance of the two phenomena, and some kind of
multi-dimensional analysis would be required.  We have proposed a
method for discriminating between these two end-of-energy models by
reconstructing the individual emission spectra from various nearby
sources.

We proposed to combine information on nuclear composition and arrival
direction to associate a potential clustering of trans-GZK events with
a 3-dimensional position in the sky. For a given cluster, the distance
to the source and its maximum energy could be determined through a
multi-parameter fit to the observed nuclear composition of each
individual event, in conjunction with possible GZK energy losses. This
allows for a model discrimination on an statistical basis by comparing
the maximum energy at the source of each individual cluster. 

We have identified a striking difference between the anisotropy
patterns created by proton- and nucleus-emitting-sources. On the one
hand, sources of UHECR nuclei display (after CR propagation)
anisotropy patterns in the shape of onion layers, with radii that
increase with rising energy. These hot spots are expected to shine in
the sky at energies $10^{10.6} \alt E/{\rm GeV} \alt 10^{11}$. A prime
example of sources of UHECR nuclei are starbursts
galaxies~\cite{Anchordoqui:1999cu}. The Pierre Auger Collaboration has
recently reported an indication of a possible correlation between CR
with $E > 3.9 \times 10^{10}~{\rm GeV}$ and starburst galaxies, with
an {\it a posteriori} significance level of
$4\sigma$~\cite{Aab:2017njo,ObservatoryMichaelUngerforthePierreAuger:2017fhr}.
The smearing angle and the anisotropic fraction corresponding to the
best-fit parameters are $13^\circ$ and 10\%.  On the other hand,
sources of UHECR protons display anisotropy patterns which become
denser and compressed with rising energy. The Pierre Auger
Collaboration has also reported a less significant ($2.7\sigma$)
correlation between CR with $E > 6.0 \times 10^{10}~{\rm GeV}$ and the
brightest radio loud active galactic nuclei (within a
250~Mpc radius from Earth) from the second catalog of hard {\it Fermi}-LAT
sources~\cite{Aab:2017njo,ObservatoryMichaelUngerforthePierreAuger:2017fhr}. The
smearing angle and the anisotropic fraction corresponding to the
best-fit parameters are $7^\circ$ and 7\%. The high energy threshold
and distinctively the smaller size of the hot spots may be indicative
of UHECR protons. If this were the case these sources will be uncover
in the very near future. This is because the study of UHECRs with
POEMMA will yield orders-of-magnitude increase in statistics of
observed UHECRs, particularly beyond $10^{11}~{\rm GeV}$ where proton
sources become unmistakable.

A point worth noting at this juncture is that albeit at first glance
the analysis technique and conclusions presented herein may appear
similar to those of~\cite{Allard:2008gj}, there are a few key differences
which make the analyses complementary to one
another. In~\cite{Allard:2008gj} the authors consider all UHECR data with
equal weights, and study the spectral shape fitting simultaneously both  sub- and
trans-GZK events. This study provides a general description of the observed
UHECR spectrum, but ignores highly technical details of the nuclear
composition at the high
energy end of the spectrum. For example, the authors
of~\cite{Allard:2008gj} conclude that {\it light and
  intermediate (He, CNO and Si; respectively) mass nuclei are not expected to play any
  significant role above $\sim 10^{10.7}~{\rm GeV}$ due to their
  interaction with the photon backgrounds even if they were present or
  even dominant at the sources.} As we have shown in this paper, the multi-dimensional
reconstruction of the individual  trans-GZK spectra (in $E$,
direction, and cross-correlation with nearby putative sources)
highlights the importance of  these nuclear species  to uncover the {\it cepa
  stratis} structure portrayed by UHECR nucleus sources from our
cosmic backyard.

We have also revisited the hypothesis that UHECR nuclei can originate
in starburst galaxies. We have shown that iron nuclei can be stripped
off the surface of young neutron stars (which are abundant in the
central region of these galaxies) and can be accelerated without
suffering catastrophic interactions in a two-step process that
involves a one-shot acceleration in the potential drop of the
fast-spinning pulsar, followed by re-acceleration at the large scale
terminal shock produced by the superwind that flows from the starburst
engine. The acceleration process yields a hard emission spectrum
$\propto E^{-1}$, as needed to reproduce Auger data on both the
spectrum and the observed nuclear composition. When the hard source
spectrum is combined with the evidence for correlation we conclude
that {\it starburst galaxies provide a compelling source example of
  UHECRs}. Using starbursts as a working example we demonstrated the
functionality of the proposed multidimensional analysis.

In summary, we have demonstrated that our local universe encompasses a
natural mass spectrometer that can be operated to untangle the origin
of UHECRs. We have shown that while sources of nuclei design a {\it
  cepa stratis} structure in the sky, with the layer of the onion
increasing with rising energy, proton accelerators imprint hot spots
which become denser and compressed with rising energy. Future data
from POEMMA~\cite{Olinto:2017xbi} (and its pathfinder
mission~\cite{Adams:2017fjh}), TA, and AugerPrime~\cite{Aab:2016vlz}
will provide a final verdict for the ideas presented and discussed in
this paper.

 \section*{Dedication}
We dedicate this work to the memory of Haim Goldberg (1939-2017). 
We cherish our long friendships and research with Haim. 
He was brilliant, but humble, always cheerful and thoughtful. 
He inspired us with his physics insights. 
The genesis of the idea explored here is a 2002 publication of Haim Goldberg,
Diego Torres, and Luis Anchordoqui. 
    
\acknowledgments{We would like to acknowledge
  many useful discussions with our colleagues of the Pierre Auger and POEMMA
  collaborations. We thank Gerald Cecil and Leonardo
  Orazi for permission to reproduce Figs.~\ref{fig:1} and \ref{fig:2}.
  LAA is supported by the U.S. National Science Foundation (NSF)
  Grant No. PHY-1620661 and by the National Aeronautics and Space
  Administration (NASA) Grant No. NNX13AH52G. VB is supported by the
  U. S. Department of Energy (DoE) Grant No. DE-FG-02-95ER40896. TJW
  is supported by DoE Grant No. DESC-0011981. }

\end{document}